\documentclass{aa}  
\usepackage{graphicx}
\usepackage{txfonts}
\usepackage{lipsum}
\newcommand\identity{1\kern-0.25em\text{l}}

\usepackage{natbib}
\bibpunct{(}{)}{;}{a}{}{,} 

\usepackage{hyperref}

\usepackage{float}

\def\Halpha{\mbox{H\hspace{0.1ex}$\alpha$}}
\def\Hbeta{\mbox{H\hspace{0.1ex}$\beta$}}
\def\Hgamma{\mbox{H\hspace{0.1ex}$\gamma$}}
\def\Hdelta{\mbox{H\hspace{0.1ex}$\delta$}}
\def\Hepsilon{\mbox{H\hspace{0.1ex}$\varepsilon$}}

\def\Lyalpha{\mbox{Ly-\hspace{0.1ex}$\alpha$}}
\def\Lybeta{\mbox{Ly-\hspace{0.1ex}$\beta$}}

\def\CaH{\ion{Ca}{ii}~H}

\def\CaIIR{\ion{Ca}{ii}~854.2~nm}

\def\MgIIHandK{\ion{Mg}{ii}~h\,\&\,k}

\def\CIIone{\ion{C}{ii} 133.4~nm}
\def\CIItwo{\ion{C}{ii} 133.5~nm}

\def\SiIV{\ion{Si}{iv}}

\defcitealias{Krikova2023}{Paper~I}
\defcitealias{Krikova2024}{Paper~II}

\begin{document}

   \title{Small-scale energetic phenomena in \Hepsilon: Ellerman bombs, UV bursts, and small flares}

   \author{K. Krikova\inst{1,2}, T. M. D. Pereira \inst{1,2}} 

   \institute{Rosseland Centre for Solar Physics, University of Oslo, PO Box 1029 Blindern, 0315 Oslo, Norway \\
   \email{kilian.krikova@astro.uio.no}
    \and
    Institute of Theoretical Astrophysics, University of Oslo, PO Box 1029, Blindern 0315, Oslo, Norway}


 
  \abstract
   {} 
   {We investigated the potential of using \Hepsilon\ to diagnose small-scale energetic phenomena such as Ellerman bombs, UV bursts, and small-scale flares. Our focus is to understand the formation of the line and how to use its properties to get insight into the dynamics of small-scale energetic phenomena.}
   {We carried out a forward modeling study, combining simulations and detailed radiative transfer calculations. The 3D radiative magnetohydrodynamic simulations were run with the \emph{Bifrost} code and included energetic phenomena. We employed a Markovian framework to study the \Hepsilon\ multilevel source function, used relative contribution functions to identify its formation regions, and correlated the properties of synthetic spectra with atmospheric parameters.}
   {Ellerman bombs are predominantly optically thick in \Hepsilon, appearing as well-defined structures. UV bursts and small flares are partially optically thin and give rise to diffuse structures. The \Hepsilon\ line serves as a good velocity diagnostic for small-scale heating events in the lower chromosphere. However, its emission strength is a poor indicator of temperature, and its line width offers limited utility due to the interplay of various broadening mechanisms. Compared to \Halpha, \Hepsilon\ exhibits greater sensitivity to phenomena such as Ellerman bombs, as its line core experiences higher extinction than the \Halpha\ wing.}
   {\Hepsilon\  is a valuable tool for studying small-scale energetic phenomena in the lower chromosphere. It provides more reliable estimates of velocities than those extracted from wing emission in \Halpha\ or \Hbeta. Maps of \Hepsilon\ emission show more abundant energetic events than the \Halpha\ counterpart. Our findings highlight \Hepsilon's potential to advance our understanding of dynamic processes in the solar atmosphere.}

   \keywords{ Radiative transfer --
              Line: formation --
              Sun: atmosphere --
              Sun: activity --
              Sun: flares
               }

   \maketitle

\nolinenumbers
\section{Introduction}
\label{sec:Intro}

Magnetic reconnection is widely regarded as a fundamental driver of various solar energetic phenomena, including flares, Ellerman bombs \citep{Pariat2007, BelloGonzalez2013,Rouppe2016}, and UV bursts \citep{Peter2014,Young2018}.
Flares represent the most dramatic releases of magnetic energy in the solar atmosphere, with observable signatures across spectral lines originating from the corona, chromosphere, and down to the photosphere. 
UV bursts \citep{Peter2014} are chromospheric phenomena that were first observed in the \SiIV\ lines using the Interface Region Imaging Spectrograph \citep[IRIS;][]{DePontieu2014}, and they are several orders of magnitude less energetic than flares.
The smallest detectable transient reconnection events are commonly referred to as Ellerman bombs \citep{Ellerman1917, McMath1960, Severny1956}, which are observed in the solar photosphere.
Ellerman bombs are traditionally detected in regions of strong magnetic activity as intense brightenings of the extended wings of the \Halpha\ line, while they remain undetectable in the \Halpha\ line core.

\citet{Rouppe2016} show that Ellerman-bomb-like events seen in \Halpha\ are much more common than previously thought and not limited to active regions; they have even been observed in the quiet Sun, and such events have been termed quiet-Sun Ellerman bombs.
\citet{Nelson2017} and \citet{Shetye2018} further explored this phenomenon.
More recently, \citet{Joshi2020} and \citet{Joshi2022} revealed that quiet-Sun Ellerman bombs are more prevalent in \Hbeta\ than in \Halpha\ across the quiet Sun, which has sparked growing interest in using higher-order Balmer series lines as diagnostic tools for these small-scale energetic events.

The Balmer series line \Hepsilon\ is one such candidate.
In the solar spectrum, this line is only a weak blend in the \CaH\ wing.
Its formation in the solar spectrum was first studied by \citet{Ayres1975}, who used a 1D plane-parallel static model atmosphere.
In \citet[hereafter \citetalias{Krikova2023}]{Krikova2023} we revisited the formation of \Hepsilon\ using state-of-the-art 3D radiative magnetohydrodynamic simulations \citep{Carlsson2016} coupled with modern nonlocal thermodynamic equilibrium (non-LTE) radiative transfer codes \citep{Uitenbroek2001, Pereira2015}.
In \citetalias{Krikova2023} we find that most locations in the quiet Sun are optically thin to \Hepsilon\ radiation; the exceptions are regions of enhanced temperature in the lower atmosphere, where \Hepsilon\ extinction increases. 

Earlier studies of \Hepsilon\ in solar energetic phenomena \citep[e.g.,][]{Rolli1995, Rolli1998a, Rolli1998b} focused on its behavior during flares.
Advances in instrumentation, in particular the CHROMIS instrument \citep{Scharmer2006, Scharmer2017} at the Swedish 1-m Solar Telescope \cite[SST;][]{Scharmer2003}, allow us to observe \Hepsilon\ at much higher spatial resolution and have re-sparked interest in this line.
\citet{Rouppe2024} made use of such novel observations to demonstrate that \Hepsilon\ is uniquely suited for investigations of energetic phenomena such as Ellerman bombs, even more than \Halpha\ or \Hbeta.
Furthermore, \cite{Anan2024} emphasize that the polarization of \Hepsilon\ serves as a diagnostic for electric fields linked to magnetic diffusion, potentially tied to the release of magnetic energy in small-scale energetic phenomena.

Energetic events in \Hepsilon\ are often associated with emission, either in the line core or in the wings.
The \Hepsilon\ emission profiles we studied in \citetalias{Krikova2023} came from a relatively quiet 3D simulation, which lacked reconnection-driven large events, and therefore were associated with shocks.
However, given the vast observational evidence \citep{Rouppe2024} that \Hepsilon\ emission is associated with Ellerman bombs and other events, it is worth revisiting the subject using models that actually simulate these phenomena.

In this work we investigated the fundamental formation properties of \Hepsilon\ in small-scale energetic phenomena, such as Ellerman bombs, UV bursts, and small-scale flares, using active simulations \citep{Hansteen2017, Hansteen2019} run with the \emph{Bifrost} code \citep{Gudiksen2011}.
We sought to identify what atmospheric information can be derived from the \Hepsilon\ profiles for energetic events and determine where the \Hepsilon\ radiation originates. 

We explored which transitions are responsible for the observed line photons in energetic events, encoded within the multilevel source function, using the multilevel source function description we developed in \citet[hereafter \citetalias{Krikova2024}]{Krikova2024}.
Finally, we aimed to quantify whether \Hepsilon\ is formed under optically thin or optically thick conditions during small-scale energetic events, which will provide insight into how these structures appear in observations. 

The outline of the paper is as follows.
Section \ref{sec:Methods} introduces the Bifrost simulations, details the calculation of synthetic spectra, and describes the methodologies used to analyze the \Hepsilon\ spectral profiles.
In Sect. \ref{sec:Results} we provide an overview of the synthetic spectra derived from the Bifrost simulations, determining whether the line formation is optically thick or thin.
Additionally, we performed a detailed analysis of an Ellerman bomb, a UV burst, and a small-scale flare.
Section \ref{sec: General properties} investigates the relationship between line parameters and atmospheric conditions, examines the broadening mechanisms affecting \Hepsilon, and compares synthetic \Halpha\ images with \Hepsilon.
Our findings are discussed in Sect. \ref{sec:Discussion}, which is followed by concluding remarks in Sect. \ref{sec:Conclusion}.

\section{Methods}
\label{sec:Methods}

\subsection{Simulations}

We made use of two different 3D radiative magnetohydrodynamic simulations run with the \emph{Bifrost} code \citep{Gudiksen2011}.
They were solved on a Cartesian grid with $24\times 24$ Mm$^2$ in the horizontal direction and span the upper convection zone to the lower corona. 

The first simulation is described by \citet{Hansteen2017}, and we refer to it as the \texttt{cbh} simulation.
It has a horizontal grid size of 48~km, and a vertical grid size of about 20~km in the photosphere and chromosphere.
The simulation was evolved starting with a weak and uniform magnetic field, and then a horizontal magnetic flux sheet with 336~mT (3360~G) oriented in the $y$ direction was injected at the bottom boundary, which later emerges to the photosphere.
As they rise through the atmosphere, magnetic flux elements generate several heating events, several of which resemble Ellerman bombs and UV bursts, and even some small nano- or micro-flares (with temperatures of $\approx 1$~MK).
We used a single snapshot of this simulation, at a simulated time of 8200~s, in which several energetic events are seen.

The second simulation is described by \citet{Hansteen2019}, and we refer to it as the \texttt{en} simulation.
Its spatial resolution is slightly higher.
The horizontal grid size is 31.25~km, and its vertical grid size is about 12~km in the photosphere and chromosphere.
It was started from the \emph{Bifrost} simulation of \citet{Carlsson2016}, which has two main magnetic polarities at the surface.
In a similar approach to the \texttt{cbh} simulation, a horizontal magnetic flux sheet with 200~mT (2000~G) oriented in the $y$ direction was injected at the bottom boundary and allowed to evolve.
On rising and interacting with the photospheric convection and upper atmospheres, the emerging magnetic flux sheet leads to several reconnection events that resemble Ellerman bombs and UV bursts.
We used one snapshot, chosen at a simulation time of 9100~s, in which several energetic events are visible.

\subsection{Synthetic spectra}
\label{sec:Synthetic spectra}

To obtain synthetic spectra for \Hepsilon\ we followed the same procedure as in \citetalias{Krikova2023}.
The \Hepsilon\ spectral line is a weak blend in the wing of the much stronger \CaH\ line, and therefore we needed to perform radiative transfer calculations in non-LTE (outside LTE, or local thermodynamical equilibrium) for both the hydrogen and calcium atoms.

For hydrogen, we used a model atom with eight levels plus the \ion{H}{ii} continuum and incorporated line blends in the Balmer continuum radiation.
All hydrogen lines were treated using complete redistribution (CRD).
We adopted the convention where the hydrogen ground level is $n=1$ and the continuum is $n=9$.

For calcium, we used a model atom with five levels of \ion{Ca}{ii} plus the \ion{Ca}{iii} continuum.
To correctly reproduce the wing profile around \Hepsilon, we treated the \CaH\ line with partial redistribution (PRD).
All other lines \ion{Ca}{ii} were treated under CRD, to save computing time, since their indirect effect in the \CaH\ line was negligible.

For the spectral synthesis, our main workhorse was the \mbox{\emph{RH 1.5D}} code \citep{Pereira2015, Uitenbroek2001}.
This code uses the 1.5D approximation, treating each simulation column as an independent 1D plane-parallel atmosphere.
This is a reasonable assumption for the \Hepsilon\ line \citepalias[see][]{Krikova2023}, and allows more tractable run times, since 3D non-LTE calculations with PRD are extremely expensive \citep{Sukhorukov2017}. 

We wanted to also compare radiation from \Hepsilon\  with that from \Halpha, and for this line the 1.5D approximation works poorly \citep{Leenaarts2012}.
We therefore also used the \emph{Multi3D} code \citep{Leenaarts2009} to obtain \Halpha\ line profiles.
Here we used a simplified hydrogen model atom with a five-level plus continuum.
The \Lyalpha\ and \Lybeta\ lines were modeled under CRD, using Gaussian line profiles to approximate the effects of PRD, as described by \citet{Leenaarts2012}.

Finally, we made use of the \emph{Muspel} library \citep{Muspel2024} to synthesize line profiles at arbitrary inclinations.
\emph{Muspel} includes the same continuum extinction, line profile physics, and formal solvers as \mbox{\emph{RH 1.5D}}, and enables a flexible framework to load populations from \mbox{\emph{RH 1.5D}} and compute radiation for different ray inclinations.

\subsection{Relative contribution functions}

To quantify the contribution of each layer in the simulation to the \Hepsilon\ line intensity, we employed relative contribution functions instead of the traditional contribution functions for intensity \citep{Carlsson1997}.
Contribution functions tell us where in the atmosphere the line is being formed.
Relative contribution functions \citep[introduced by][to study weak blends in strong continua]{Magain1986} can distinguish between contributions to line emission and to line absorption.
We defined the relative contribution function as
\begin{equation}\label{eq:relative_contr}
     C_\mathrm{R}(\nu, z) = \frac{\textrm{d}I_\mathrm{R}}{dz} = \chi^\mathrm{l}_\nu \, \left(1- \frac{S^\mathrm{l}_\nu}{I^\mathrm{b}_\nu} \right) \,  e^{-\tau^\mathrm{R}_\nu},
\end{equation}
where $\chi^\mathrm{l}_\nu$ is the line extinction, $S^\mathrm{l}_\nu$ the line source function, $I^\mathrm{b}_\nu$ the height-dependent background (continuum) emergent intensity, and $\tau^\mathrm{R}_\nu$ the relative optical depth.
$I^\mathrm{b}_\nu$ is computed by solving the radiative transfer equation using the background source function and extinction for a ray starting at the lower boundary and up to each height point.
The relative optical depth can be calculated through cumulative integration of the relative extinction $\chi^\mathrm{R}_\nu$ expressed as
\begin{equation}\label{eq:relative extinciton}
    \chi^\mathrm{R}_\nu = \chi^\mathrm{l}_\nu + \chi^\mathrm{b}_\nu \, \frac{S^\mathrm{b}_\nu}{I^\mathrm{b}_\nu},
\end{equation}
where $\chi^\mathrm{l}_\nu$ refers to the line extinction, $\chi^\mathrm{b}_\nu$ to the background extinction, and $S^\mathrm{b}_\nu$ to the background source function.
In the case of \Hepsilon\, the background extinction and source function include both the \CaH\ line and all continuum processes.

From Eq.~(\ref{eq:relative_contr}) there are two cases when the line is absent and the relative contribution function is zero: when $\chi^\mathrm{l}_\nu = 0$ (no absorbers), or when $S^\mathrm{l}_\nu = I^\mathrm{b}_\nu$.
When there is a spectral line, atmospheric layers contribute to either relative line depression ($S^\mathrm{l}_\nu < I^\mathrm{b}_\nu$) or emission ($S^\mathrm{l}_\nu > I^\mathrm{b}_\nu$)
We give further details on the application of relative contribution functions and \Hepsilon\ in \citetalias{Krikova2023}.
For locations with \Hepsilon\ emission, using either relative contribution functions or traditional contribution functions yields similar results. 

\subsection{Multilevel source function}
\label{sec:Radiative transfer}

To quantify the physical process contributing the most to the observed line photons, encoded in the line source function $S^\mathrm{l}_\nu$, we used the multilevel source function approach (\citealt{Jefferies1968}, \citealt{Canfield1971}, \citealt{Rutten2021}, \citetalias{Krikova2024}).
The source function is divided into three different physical mechanisms responsible for adding line photons: scattering, thermal, and interlocking:
\begin{equation}
        S^l_{\nu_0} = \sigma \, \overline{J}_{\nu_0}  + \epsilon \, B_{\nu_0} (T_\mathrm{e}) + \eta \, B_{\nu_0} (T^{\star}) \label{eq:msf_sf}.
\end{equation}
The terms $\overline{J}_{\nu_0}$, $B_{\nu_0} (T_\mathrm{e})$, and $B_{\nu_0} (T^{\star})$ represent the sources of photons related to the mean radiation field, the Planck function, and the interlocking source function, respectively. 
The coefficients $\sigma$, $\epsilon$, and $\eta$ describe which of these three physical mechanisms predominates in contributing to the line photons.

To compute the terms in Eq. (\ref{eq:msf_sf}), we used a Markovian description of the multilevel source function \citepalias{Krikova2024}. 
Additionally, we simplified the source function by assuming CRD and neglecting the stimulated emission term, which has a minimal effect in the shorter wavelength region of the solar spectrum. 
This leads to the following expression:
\begin{equation}\label{eq:source_function_CRD}
    S^l_{\nu_0} = \frac{2 h \nu_0^3}{c^2} \frac{1}{\frac{g_u n_l}{g_l n_u} - 1} \approx  \frac{2 h \nu_0^3}{c^2} \frac{g_l n_u}{g_u n_l}, 
\end{equation}
where $n_i$ and $g_i$ are the populations and statistical weights of level $i$ ($u$ for upper level and $l$ for lower level). 
Using a general solution of the statistical equilibrium equation in terms of a level-ratio solution, as expressed in \citetalias{Krikova2024}, we have
\begin{equation}\label{eq:se_solution}
     \frac{n_u}{n_l} = \frac{P_{lu} + \sum_l}{P_{ul} + \sum_u} =  \frac{P_{lu} + \sum_{i \neq u} P_{li} \, q_{iu,l}}{P_{ul} + \sum_{i \neq l} P_{ui} \, q_{il,u}}.
\end{equation}
$P_{ul}$ and $P_{lu}$ are the direct transition rates (radiative and collisional) between the upper and lower levels, and vice versa.
$\sum_u$ and $\sum_l$ contain the indirect transitions per second between the upper and lower levels through all intermediate levels.
We replaced this population ratio in Eq.~(\ref{eq:source_function_CRD}), obtaining
\begin{equation}\label{eq:source_function_markov}
    S^l_{\nu_0} = \frac{2 h \nu_0^3}{c^2} \frac{g_l}{g_u} \frac{ \left( P_{lu} + \sum_{i \neq u} P_{li} \, q_{iu,l} \right) }{ \left( P_{ul} + \sum_{i \neq l} P_{ui} \, q_{il,u} \right) }, 
\end{equation}
with the indirect transition rates split into individual intermediate levels,  $i$, with $P_{li}$  and  $P_{ui}$  representing the transition rates from the lower or upper levels to the intermediate level. 
The indirect transition probabilities, $q_{il,u}$ and $q_{iu,l}$, describe the likelihood that a transition from the intermediate level  $i$  ends up in the lower or upper level, respectively. 
The algebraic expressions for these probabilities can be quite lengthy, but in \citetalias{Krikova2024} we identify a pattern in these expressions for the indirect transition probabilities.
We utilized these patterns, which resemble all nonrecurrent paths from the intermediate level  $i$ to the lower or upper level (corrected for closed loops), to identify the dominant path that determines the level ratio and thus the line source function, as given in Eq. (\ref{eq:source_function_markov}).

\begin{figure*}
    \centering
    \includegraphics[width=1\textwidth]{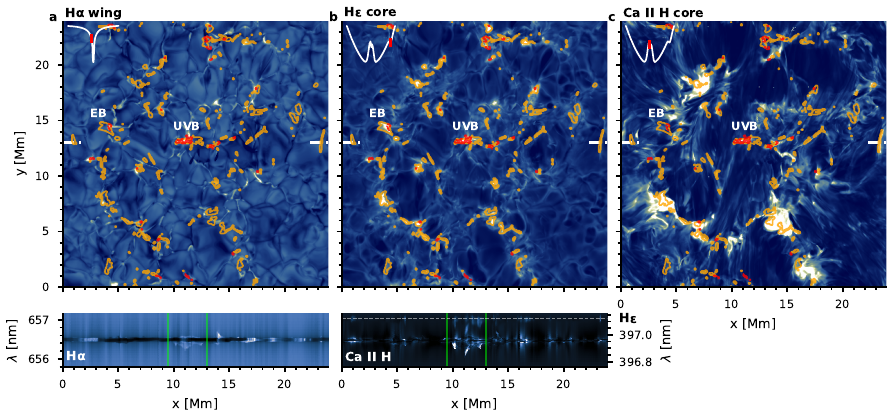}
    \caption{Synthetic observables from the \texttt{en} simulation. The top three panels show images at the fixed wavelengths in the \Halpha\ wing, \Hepsilon\ core, and \CaH\ core, respectively. The bottom two panels show spectrograms normalized to the continuum taken from a horizontal slice in the simulation at $y\approx 13$~Mm (edges denoted by the white lines in the top panels), in \Halpha\ and \CaH\ (which also includes \Hepsilon, whose location is indicated by the dashed gray line). The orange contours in the top panels indicate regions where \Hepsilon\ is strongly in emission, while the red contours indicate regions where \Halpha\ shows the typical signature of Ellerman bombs. The UVB and EB labels indicate the location of a UV burst and an Ellerman bomb selected for detailed study. The green vertical lines in the bottom panels indicate the approximate extent of the UV burst.}
    \label{fig:en_overview}
\end{figure*}

\subsection{Optical depth of \Hepsilon}

We want to quantify whether the structures observed in \Hepsilon\ are formed under optically thin or thick conditions and follow an approach similar to that of \citet{Rathore2015}, who introduce a mean formation depth $\tau_{\mathrm{fm}}$ (Eq. 4), in terms of $\log_{10}\tau_0$ (the optical depth at the line core), to quantify the significance of optically thick or thin formation in the \CIIone\ and \CIItwo\ lines.

We introduced an average formation optical depth $\tau_{\mathrm{afd}}$, defined as%
\begin{gather}
     \log_{10}\tau^{\mathrm{fd}}_{\lambda} = \frac{\int \log_{10}\tau_\lambda \: |C_{R_\lambda}(\log_{10} \tau_\lambda)| \: \mathrm{d}\log_{10}\tau_\lambda}{ \int |C_{R_\lambda}(\log_{10}\tau_\lambda)| \: \mathrm{d}\log_{10}\tau_\lambda},
     \\
     \log_{10}\tau_{\mathrm{afd}} = \frac{1}{\lambda_{\mathrm{max}} -\lambda_{\mathrm{min}}} \int_{\lambda_{\mathrm{min}}}^{\lambda_{\mathrm{max}}} \log_{10} \tau^{\mathrm{fd}}_{\lambda} \: \mathrm{d}\lambda \label{eq:afd},
\end{gather}
where $\log_{10}\tau^{\mathrm{fd}}_{\lambda}$ quantifies the optical thickness or thinness at each wavelength position across the spectral profile on a logarithmic optical depth scale.
The quantity $\tau_{\mathrm{afd}}$ represents the average optical thickness or thinness of structures within a spectral line, calculated over the wavelength range bounded by $\lambda_{\mathrm{max}}$ and $\lambda_{\mathrm{min}}$.  
The greater the deviation from zero (i.e., the more negative $\tau_{\mathrm{afd}}$ becomes), the more optically thin structures become in a spectral line. 

\subsection{Line parameters and atmospheric conditions}
\label{sec:line_parameters}

During energetic events, \Hepsilon\ is often in emission.
To quantify the properties of these emission profiles, we employed a composite fitting model comprising a Gaussian component and a second-degree polynomial background for the synthetic \Hepsilon\ profiles.
Such fitting is always approximate since the line shapes vary considerably, but we find it a good approximation to extract the line shift, width, and peak intensity.
The second-degree polynomial effectively accounts for the variability in the local continuum of the \CaH\ wing.
The resulting \Hepsilon\ line profile is then approximated as
\begin{equation}\label{eq:gaussian}
    \phi(\lambda) = \frac{1}{\sqrt{\pi} \Delta \lambda_\mathrm{D}^\mathrm{fit}} \exp{-\left(\frac{\lambda - \lambda_0}{\Delta \lambda_\mathrm{D}^\mathrm{fit}}\right)^2}, 
\end{equation}
where $\lambda_0$ is the fitted line shift and $\Delta \lambda_\mathrm{D}^\mathrm{fit}$ the line width. 

For optically thin lines, the Doppler width is a combination of thermal and nonthermal broadening:
\begin{equation}\label{eq:theoretical_broadening}
    \Delta \lambda_\mathrm{D} = \sqrt{\frac{2 k_\mathrm{B} T}{m} + \xi^2}, 
\end{equation}
where $\xi$ describes the nonthermal broadening. Following the approach of \citet{Rathore2015}, we estimated the nonthermal broadening from the root mean square of the line of sight velocities extracted from the simulation, in the line-forming region:
\begin{equation}\label{eq:nonthermal_velocity}
    \xi = \sqrt{\frac{1}{n} \sum_{i=\mathrm{h_{cont}}}^{\mathrm{h_{rw}}} \mathrm{v}_i^2}, 
\end{equation}
where $\mathrm{v}_i$ is the line of sight velocity at height $i$, $\mathrm{h_{cont}}$ is the height where the background is formed (we assumed at $\lambda = 396.85$~nm, the symmetric blue wing position of \Hepsilon\ at the \CaH\ wing) and $\mathrm{h_{rw}}$ is relative contribution function weighted average height.

Why we used $\mathrm{h_{rw}}$ to estimate the nonthermal broadening and how we identified the formation region of \Hepsilon\ warrants further explanation. When a spectral line is optically thick, $h(\tau_\lambda=1)$, the height where the optical depth reaches unity, is a good proxy for the height of formation. Indeed, in several forward-modeling studies, authors correlate atmospheric properties at $h(\tau_\lambda=1)$ with parameters of spectra \citep[e.g.,][]{Leenaarts2013, Pereira2013, Pereira2015a, Rathore2015, Leenaarts2016, Lin2018}. 
However, when a line is formed under optically thin conditions, or if its source function varies nonlinearly over the line formation region, $h(\tau_\lambda=1)$ is not a reliable proxy.
A more accurate approach to obtaining estimates of height, temperature, velocity, or other atmospheric parameters at the region where the line forms is to weigh the quantity $X$ by the relative contribution function (Eq. \ref{eq:relative_contr}):
\begin{equation}\label{eq:relative_cf_weighted}
    X_\mathrm{rw} = \frac{\int_0^\infty C_\mathrm{R}(\nu, z) \, X(z) \mathrm{d}z}{\int_0^\infty C_\mathrm{R}(\nu, z) \mathrm{d}z}. 
\end{equation}

\noindent We used this definition to obtain not only $\mathrm{h_{rw}}$ and $\xi$, but also temperature, line of sight velocity, and thermal broadening.

\citet{Rathore2015b} demonstrated that the optically thick \CIIone\ and \CIItwo\ lines can be significantly affected by opacity broadening.
Opacity broadening occurs when the observed spectral line width exceeds what can be explained solely by Doppler broadening (thermal and nonthermal contributions), with the excess width arising from opacity effects in an optically thick medium.
To quantify opacity broadening effects in \Hepsilon, we studied how the fitted line width $\Delta \lambda_\mathrm{D}^\mathrm{fit}$ follows the theoretical $\Delta \lambda_\mathrm{D}$ from Eq.~(\ref{eq:theoretical_broadening}). 
We built $\Delta \lambda_\mathrm{D}$ by using $\xi$ from Eq.~(\ref{eq:nonthermal_velocity}) and the thermal velocity
\begin{equation}\label{eq:vthermal}
    v_\mathrm{th} = \sqrt{\frac{2k_B T_\mathrm{rw}}{m}}, 
\end{equation}
where $T_\mathrm{rw}$ is the temperature in the line-forming region following Eq.~(\ref{eq:relative_cf_weighted}). Therefore, $\Delta \lambda_\mathrm{D}$ depends only on quantities from the simulation, while $\Delta \lambda_\mathrm{D}^\mathrm{fit}$ comes from fitting a Gaussian to the synthetic profile. In the absence of opacity broadening these should be similar, and we defined an opacity broadening factor $Opf$ as
\begin{equation}\label{eq:opf}
    Opf = \frac{\Delta \lambda_\mathrm{D}^\mathrm{fit}}{\Delta \lambda_\mathrm{D}}.
\end{equation}

If Eq.~(\ref{eq:theoretical_broadening}) holds for the synthetic profiles, we can use it to measure the nonthermal broadening from the spectra (as opposed to $\xi$, which is obtained from the atmospheric velocities), using $v_\mathrm{th}$ and $\Delta \lambda_\mathrm{D}^\mathrm{fit}$:
\begin{equation}\label{eq:nonthermal_broadening}
    v_\mathrm{nth} =  \sqrt{\left(\Delta \lambda_\mathrm{D}^\mathrm{fit}\right)^2 - v_\mathrm{th}^2}.
\end{equation}

\noindent In the absence of opacity broadening, and assuming a Gaussian profile, $v_\mathrm{nth}$ should closely follow $\xi$.

\section{\Hepsilon\ in energetic events}
\label{sec:Results}

\subsection{Synthetic observables}
\label{sec: Synthetic observables}

\begin{figure*}
    \centering
    \includegraphics[width=1\textwidth]{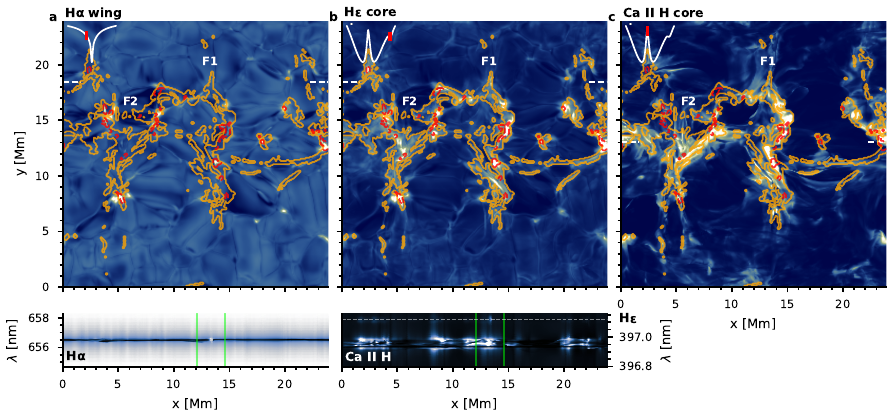}
    \caption{Synthetic observables from the \texttt{cbh} simulation. The caption is the same as for Fig. \ref{fig:en_overview} with the difference that the labeled events are now F1 and F2, the locations of two small flares. The vertical green lines in the bottom panels indicate the approximate position of the F1 flare.}
    \label{fig:cb_overview}
\end{figure*}

We present an overview of the synthetic observations of \Halpha, \Hepsilon, and \CaH\ in Fig.~\ref{fig:en_overview} for the \texttt{en} simulation, and in Fig.~\ref{fig:cb_overview} for the \texttt{cbh} simulation.
Both simulations contain Ellerman bombs and UV bursts, and \texttt{cbh} further contains two small-scale flares.

In the figures, we show the \Halpha\ intensity at a wing position, typically used to identify Ellerman bombs, which have a mustache-like profile with raised wings.
We define regions of strong \Hepsilon\ emission as where the \Hepsilon\ core intensity exceeds 5\% of its background.
The background is defined as the \CaH\ wing intensity at 396.85~nm, corresponding to the symmetric blue wing position of \Hepsilon.
As a result, not all Ellerman bombs detected in \Halpha\ necessarily coincide with regions of strong \Hepsilon\ emission.

The synthetic spectra from both simulations confirm the findings of \citetalias{Krikova2023}, in particular that \Hepsilon\ emission is an indicator of lower atmospheric heating.
From Figs.~\ref{fig:en_overview} and \ref{fig:cb_overview} we see that \Hepsilon\ emission is more common than in the quiet-Sun simulation of \citetalias{Krikova2023} and that it is often concentrated near intergranular lanes, where the magnetic field is concentrated.
\Hepsilon\ emission is often colocated with or adjacent to \Halpha\ Ellerman bomb signatures, and regions surrounding the small flares have widespread emission in \Hepsilon.
We examine a prominent Ellerman bomb, labeled in Fig. \ref{fig:en_overview}, in more detail in Sect. \ref{sec:Ellerman bomb} to determine the origin of the \Hepsilon\ emission.

\begin{figure*}
    \centering
    \includegraphics[width=1\textwidth]{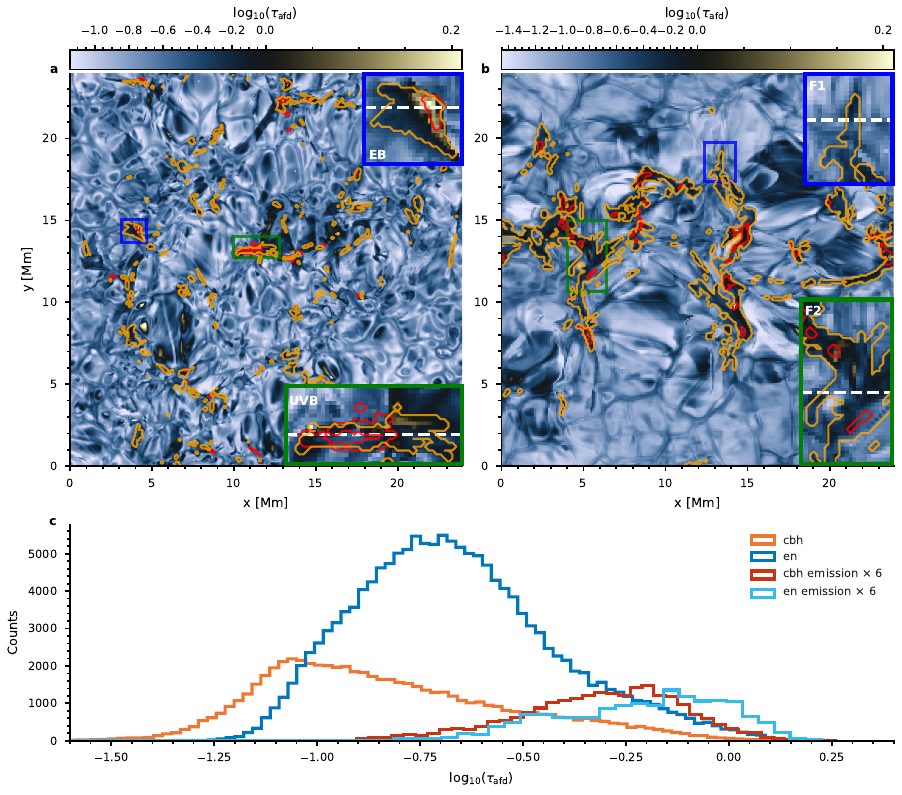}
    \caption{Average formation depth, $\tau_{\mathrm{afd}}$, for the \Hepsilon\ spectral line. \emph{Top panels:} $\tau_{\mathrm{afd}}$ maps for the \texttt{en} simulation (panel \emph{a}) and \texttt{chb} simulation (panel \emph{b}). The orange contours indicate regions where \Hepsilon\ is strongly in emission, while the red contours indicate regions where \Halpha\ shows the typical signature of Ellerman bombs. For each panel, two inserts show a zoomed-in view covering the energetic events labeled earlier: an Ellerman bomb (EB), a UV burst (UVB), and two small flares (F1 and F2). The dashed white lines in the insets indicate the horizontal slices taken for the spectrograms of Figs.~\ref{fig:en_overview} and \ref{fig:cb_overview}. \emph{Panel c:} Distributions of $\log_{10}\tau_{\mathrm{afd}}$. The emission distribution is amplified by a factor of six.}
    \label{fig:en_cbh_optick_thick_thin}
\end{figure*}

Figure \ref{fig:en_overview} also demonstrates that UV bursts can produce \Hepsilon\ emission signatures. 
This is illustrated in the \CaH\ plus \Hepsilon\ spectro-heliograms, which show multiple strong \Hepsilon\ emission locations throughout the UV burst structure. 
In Sect. \ref{sec: UV burst} we explore the source of the \Hepsilon\ intensity in a UV bursts.
Our synthetic profiles of Ellerman bombs and UV bursts (e.g., lower panels of Fig. \ref{fig:en_overview} in between the green lines) show wing emission in \Halpha\ (with little to no signature in the core) and core emission in \Hepsilon, a signature that is consistent with observations \citep{Rouppe2024}.

We also investigated the formation properties of \Hepsilon\ in small-scale flares, labeled F1 and F2 in Fig. \ref{fig:cb_overview}.
F2 is the more intense of the two.
The spectrograms show a slice across F1, where both \Halpha\ and \Hepsilon\ are in emission at the flaring location. 
In Sect. \ref{sec: Small-scale flares} we discuss the formation mechanisms of \Hepsilon\ for these small-scale flare events.

To complement the spectral information, we map $\tau_{\mathrm{afd}}$ in Fig.~\ref{fig:en_cbh_optick_thick_thin} to identify where \Hepsilon\ is optically thin and optically thick for both simulations. We find that \Hepsilon\ is optically thin in most locations in both simulations, but that regions with \Hepsilon\ emission tend to be optically thicker, as seen in the distributions in the lower panel of Fig.~\ref{fig:en_cbh_optick_thick_thin}.

The distribution for the \texttt{en} simulation peaks at $\mathrm{log}(\tau_{\mathrm{afd}}) = -0.75$, whereas the \texttt{cbh} distribution is skewed toward lower $\mathrm{log}(\tau_{\mathrm{afd}})$ values. 
The light blue and red distributions indicate that \Hepsilon\ emission regions are optically thicker compared to their surroundings (orange and blue distributions), although a significant fraction of the locations still show optically thin structures in \Hepsilon.

To gain a more detailed perspective on the energetic events, we inserted magnified views of an Ellerman bomb, a UV burst, and small-scale flares into the corners of Fig. \ref{fig:en_cbh_optick_thick_thin}.
The \Hepsilon\ emission associated with the Ellerman bomb structure is predominantly formed under optically thick conditions, with most of the contribution originating near the $\tau_\lambda=1$ layers. 
In locations where \Halpha\ and \Hepsilon\ simultaneously display Ellerman bomb signatures, it is evident that part of the \Hepsilon\ contribution arises from below the $\tau_\lambda=1$ layers. 
However, as we approach the western edge of the Ellerman bomb structure, the \Hepsilon-emitting regions increasingly display optically thin characteristics, with contributions coming from above the $\tau_\lambda=1$ layer.

The insets for more energetic events show a more dispersed distribution of optically thin and thick structures.
The UV burst inset reveals that a large fraction of the \Hepsilon\ emission associated with the UV burst structure is formed near the $\tau_\lambda=1$ height, although some parts of the burst are optically thin. 
For the small-scale flares (F1 and F2), a significant portion of the structures seen in \Hepsilon\ are optically thin, indicating that layers above $\tau_\lambda=1$ contribute predominantly to the emergent line intensity.

\subsection{Ellerman bomb}
\label{sec:Ellerman bomb}

We next looked in detail at the Ellerman bomb structure labeled in Fig. \ref{fig:en_cbh_optick_thick_thin}.
We aimed to determine which parts of the Ellerman bomb are observed in the \Hepsilon\ line and how this compares to other chromospheric lines.
Additionally, we identified the transitions that govern the \Hepsilon\ source function within the Ellerman bomb, using Eq.~(\ref{eq:source_function_markov}).

\begin{figure}
    \centering
    \includegraphics[width=0.5\textwidth]{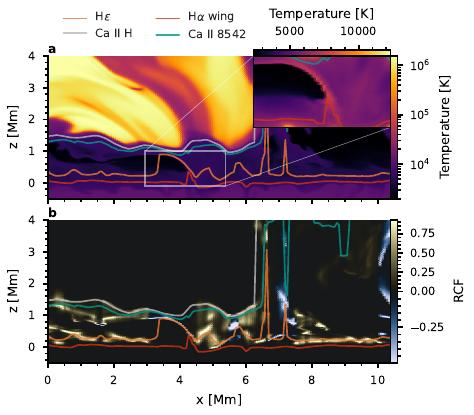}
    \caption{Line formation through an Ellerman bomb. \emph{Panel a:} Temperature slice across the atmosphere with $h(\tau_\lambda=1)$ overplotted for the different line cores of \Hepsilon, \CaH, and \CaIIR\  and the wing of \Halpha\ at $656.35$~nm. The inset shows a zoomed-in view around the Ellerman bomb, excluding $h(\tau_\lambda=1)$ of \Hepsilon; there is a different temperature scale on top. \emph{Panel b:} Relative contribution function at the maximum $h(\tau_\lambda=1)$ over the \Hepsilon\ profile.}
    \label{fig:eb_temp_rcf}
\end{figure}

In Fig.~\ref{fig:eb_temp_rcf} we present the horizontal temperature profile of the Ellerman bomb, along with the \Hepsilon\ relative contribution function, to identify the layers where most line photons originate from. 
The relative contribution function is shown for the wavelength position where the maximum $\tau_\lambda=1$ height is reached within \Hepsilon. 
In the top panel of Fig.~\ref{fig:eb_temp_rcf} we overplot the $\tau_\lambda=1$ heights for \Hepsilon\ with those of \CaH, \CaIIR, and the wing of \Halpha\ (at $656.35$~nm).
\CaH\ and \CaIIR\ probe the upper chromosphere above the Ellerman bomb structure, whereas the Ellerman bomb itself is visible in the \Halpha\ wing.
Between $x = 4$ and $x = 4.5$~Mm, we observe an increase in the $\tau_\lambda=1$ height of the \Halpha\ wing associated with the Ellerman bomb structure.

The \Halpha\ wing captures the lower part of the Ellerman bomb, where temperatures range between $7000$ and $8000$~K, as shown in the inset of panel (a).
The inset further reveals a temperature front, with values around 7500~K, extending and tilting upward.
This temperature front is observed in \Hepsilon, as evidenced by the $\tau_\lambda=1$ height, which follows closely the front.
This front represents the layer that predominantly contributes to the emergent line intensity, as emphasized by the relative contribution function in panel (b).
Panel (b) appears largely dark over the heights where the temperature front is located because the contribution to line emission (light blue color) is precisely situated beneath the $\tau_\lambda = 1$ line in the image.
Consequently, this spatial alignment between the relative contribution function and the $\tau_\lambda = 1$ height highlights optically thick line formation.
An exception occurs at the leftmost edge of the structure ($3 < \mathrm{x}\;\mathrm{(Mm)} < 3.3$), where $\tau_\lambda=1$ is reached much lower than the region from which \Hepsilon\ photons originate (light blue area).

\begin{figure}
    \centering
    \includegraphics[width=0.5\textwidth]{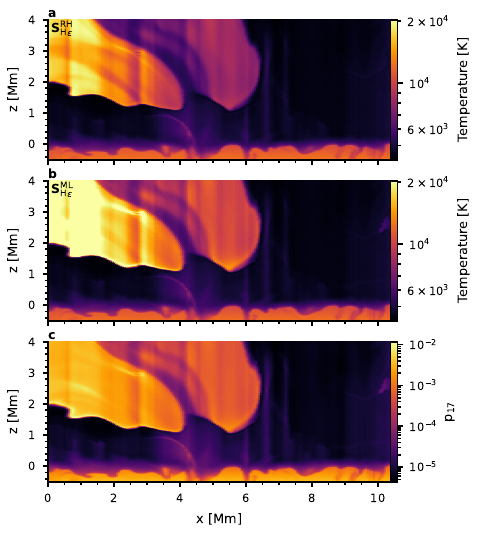}
    \caption{Source function through an Ellerman bomb. \emph{Panel a:} \Hepsilon\ source function as computed by \emph{RH 1.5D}. \emph{Panel b:}  Approximate \Hepsilon\ source function  from Eq.~(\ref{eq:source_function_markov_approx}). Both source functions are shown on a brightness temperature scale. \emph{Panel c:} Transition probability $p_{17}$ from $n=1$ to the upper level of \Hepsilon.}
    \label{fig:eb_sfrh_sfmulti}
\end{figure}

To quantify which transition pathways control the \Hepsilon\ source function, responsible for the majority of the emitted line photons, we present the multilevel source function in Fig. \ref{fig:eb_sfrh_sfmulti}.
It highlights why Ellerman bombs appear brighter than their surroundings.
The Ellerman bomb structure shows an increase in the line source function that partly resembles the temperature structure shown in Fig. \ref{fig:eb_temp_rcf}. 
In \citetalias{Krikova2024} we explored the nature of indirect transition probabilities $q_{il,u}$ and $q_{iu,l}$ and provide an analytical framework to express them.

By analyzing the transition probabilities for this event, we find that the \Hepsilon\ source function, as described by Eq. (\ref{eq:source_function_markov}), is primarily determined by transition paths involving the ground level, denoted by $P_{21} q_{17,2}$ and $P_{71} q_{12,7}$.
$P_{21}$ is the total transition rate from the lower level of the \Hepsilon\ transition to  $n = 1 $, and $q_{17,2}$ the probability of reaching the upper level of the \Hepsilon\ transition.
The total transition rate  $P_{71}$, combined with the probability $q_{12,7}$, describes the reverse transitions, originating from the upper level of \Hepsilon.

Looking further, we find that the dominant pathways into the transition probabilities $q_{17,2}$ and $q_{12,7}$ are the first-order paths, which connect directly the upper and lower levels of \Hepsilon\ with the ground level.
These first-order paths are described by the probabilities $p_{17}$ and $p_{12}$.
This means that the \Hepsilon\ source function can be approximated as
\begin{equation}\label{eq:source_function_markov_approx}
    S^\mathrm{ML}_{\nu} \propto \frac{2 h \nu_0^3}{c^2} \frac{g_l}{g_u} \frac{ \left( P_{21}  q_{17,2} \right) }{ \left(P_{71}  q_{12,7} \right)} \propto p_{17}.
\end{equation}

In Fig.~\ref{fig:eb_sfrh_sfmulti} we show $S_\nu^\mathrm{RH}$, the line source function calculated in the \emph{RH 1.5D} code, $S^\mathrm{ML}_{\nu}$ (both on a brightness temperature scale), and $p_{17}$ near the Ellerman bomb.
We present the source functions expressed in brightness temperatures, which correspond to the temperature for which the Planck function reproduces the observed intensity to facilitate direct comparison with the atmospheric temperature.
It is clear that in the Ellerman bomb region, they all show similar morphologies, strongly suggesting that the approximation in Eq.~(\ref{eq:source_function_markov_approx}) is valid.
The number of photons detected from the Ellerman bomb structure closely correlates with the transition probability $p_{17}$.

\subsection{UV burst}
\label{sec: UV burst}

\begin{figure}
    \centering
    \includegraphics[width=0.5\textwidth]{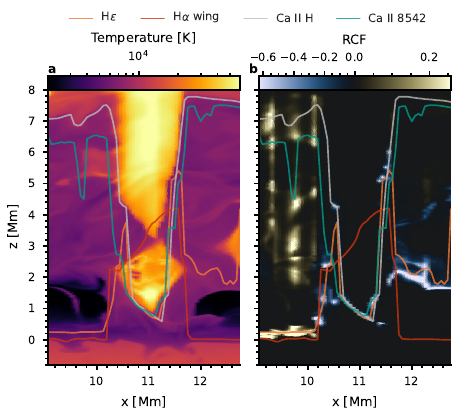}
    \caption{Line formation through a UV burst. The caption is the same as for Fig. \ref{fig:eb_temp_rcf}.}
    \label{fig:uvb_cf_rcf}
\end{figure}

We depict a detailed view of line formation for a UV burst in Fig.~\ref{fig:uvb_cf_rcf}. 
Similar to Fig.~\ref{fig:eb_temp_rcf}, it shows the temperature and \Hepsilon\ relative contribution function with $h(\tau_\lambda=1)$ for different lines. 

At this instant in the \texttt{en} simulation ($t = 8\,200$~s), the UV burst exhibits moderate temperatures above $10\,000$~K.
In the temperature cut, it appears as a lozenge shape of enhanced temperature, at heights between 1 and 3~Mm in the chromosphere.

\Hepsilon\ traces the deepest layers of the UV burst, as shown by its $h(\tau_\lambda=1)$, where it becomes optically thick to \Hepsilon\ radiation.
This is confirmed by the overlap of the relative contribution function with $h(\tau_\lambda=1)$.
The temperatures in this \Hepsilon\ formation region range between $8\,000$ and $14\,000$~K.
The \CaH\ and \CaIIR\ lines are sensitive to approximately the same UV burst layers as \Hepsilon, and thus reveal comparable structures.
In contrast, the \Halpha\ wing intensity is formed above \Hepsilon.

\begin{figure}
    \centering
    \includegraphics[width=0.5\textwidth]{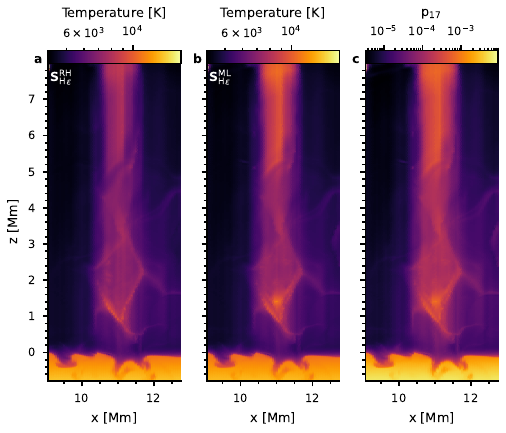}
    \caption{Source function through a UV burst. The caption is the same as for Fig.~\ref{fig:eb_sfrh_sfmulti}.}
    \label{fig:uvb_sfrh_sfmulti}
\end{figure}

Similar to the results for the Ellerman bomb, we find that the \Hepsilon\ source function in the UV burst region is dominated by a first-order transition path through the ground level, as revealed by an analysis of the multilevel source function.
We show this in Fig. \ref{fig:uvb_sfrh_sfmulti}, where we compare the line source function with the multilevel approximation from Eq.~(\ref{eq:source_function_markov_approx}) and $p_{17}$. 
In these panels, the UV burst structure is clearly visible in the transition probability $p_{17}$, with finer structures extending from the UV burst.
These finer structures, in turn, appear faintly in the temperature structure with significant contribution to relative emission, as shown in panel (b) of Fig. \ref{fig:uvb_cf_rcf}.
The most distinct of these structures is located between $x=12$ and $x=12.8$~Mm, at an approximate height of $2$~Mm, and is responsible for \Hepsilon\ emission that is not directly associated with the UV burst.

\subsection{Small-scale flares}
\label{sec: Small-scale flares}

We next turned our attention to the small-scale flares in the \texttt{cbh} simulation.
We focused on two events, F1 and F2. F1 is the less energetic of the two.
It reaches temperatures of $10\,000$~K, which is not high enough to fully ionize the chromosphere.
F2, on the other hand, reaches temperatures in excess of 1~MK.

\begin{figure}
    \centering
    \includegraphics[width=0.5\textwidth]{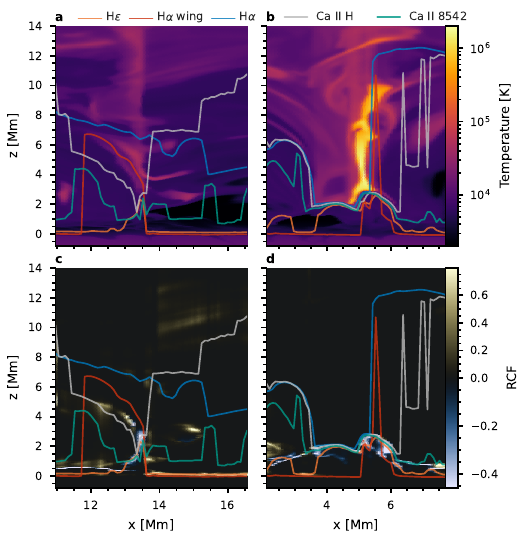}
    \caption{Line formation through two small-scale flares. The caption for each pair of panels is the same as for Fig.~\ref{fig:eb_temp_rcf}. Panels a and c are for the F1 flare, while panels b and d are for the F2 flare. In this figure there is an additional line for the $h(\tau)=1$ of \Halpha\ line core.}
    \label{fig:flare_cf_rcf}
\end{figure}

We show in Fig.~\ref{fig:flare_cf_rcf} the detailed temperature and \Hepsilon\ relative contribution function with $h(\tau_\lambda=1)$ for different lines.
This is similar to Figs.~\ref{fig:eb_temp_rcf} and \ref{fig:uvb_cf_rcf}, but a difference here is that we also show $h(\tau_\lambda=1)$ for the \Halpha\ core.

For the F1 flare, Fig.~\ref{fig:flare_cf_rcf} confirms that the chromosphere is not fully ionized, since $h(\tau_\lambda=1)$ for \Halpha\ core is formed above the flare.
\Hepsilon, the \Halpha\ wing, \CaH, and \CaIIR\ observe distinct parts of the event.
The temperature structure of the flare is best traced at the \Halpha\ wing.
\Hepsilon, \CaH, and \CaIIR\ are sensitive to the lower part of the flare's temperature enhancement, observing similar regions.
The \Hepsilon\ contribution to relative line emission comes from layers between $1$ and $2$~Mm, where temperatures reach close to $7\,000$~K.
This contribution lies slightly above $h(\tau_\lambda=1)$, particularly at $x = 13.5$~Mm, indicating that parts of the flare are optically thin to \Hepsilon\ radiation.

The much hotter F2 flare leads to a significant degree of hydrogen ionization, which results in negligible \Halpha\ extinction and an optically thin chromosphere, as observed in \Halpha.
Thus, both \Halpha\ and \Hepsilon\ observe the lower part of the flare, where temperatures remain low enough to avoid full hydrogen ionization.
While \Halpha\ traces a higher layer of the flare due to its higher line extinction per particle (about 50 times that of \Hepsilon), $h(\tau_\lambda=1)$ of \Hepsilon\ lies just below the region where most of the contribution to relative line emission originates, emphasizing that parts of the flare are optically thin to \Hepsilon\ radiation.

\begin{figure}
    \centering
    \includegraphics[width=0.5\textwidth]{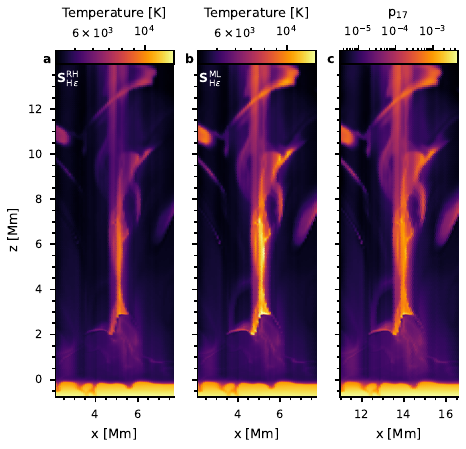}
    \caption{Source function through small-scale flare F2. The caption is the same as for Fig.~\ref{fig:eb_sfrh_sfmulti}.}
    \label{fig:flare_sfrh_sfmulti}
\end{figure}

In Fig. \ref{fig:flare_sfrh_sfmulti} we examine which term in the multilevel source function (Eq. \ref{eq:source_function_markov}) dominates the \Hepsilon\ source function for the hotter F2 flare.
Our analysis of the multilevel source function reveals that the proportionality given in Eq.~(\ref{eq:source_function_markov_approx}) holds true also for the flare structure.
The transition probability $p_{17}$ accurately represents the variation in the \Hepsilon\ source function throughout the flare, particularly in the fine structures around $2$~Mm and below.
These fine structures also highlight where the relative contribution to line emission originates from at $x = 6.15$~Mm from approximately $z = 1$~Mm and beyond.
An increase in $p_{17}$ leads to an enhanced source function and, consequently, increased \Hepsilon\ emission.

\section{General properties of \Hepsilon\ in emission}
\label{sec: General properties}
\subsection{Line properties}
\label{sec: Hepsilon line moments}

The shapes of spectral lines encode essential information about the atmospheric conditions under which the lines form.
We studied the shapes of \Hepsilon\ profiles in the regions from both simulations where they are in emission and performed Gaussian fitting to extract three key parameters: line core intensity, width, and Doppler shift.
From these, we computed additional quantities such as thermal and nonthermal broadening, and an estimate of opacity broadening (Sect.~\ref{sec:line_parameters}).
We then compared the parameters derived from the spectra with atmospheric quantities (weighted by the relative contribution function of \Hepsilon). 

\begin{figure*}
    \centering
    \includegraphics[width=1\textwidth]{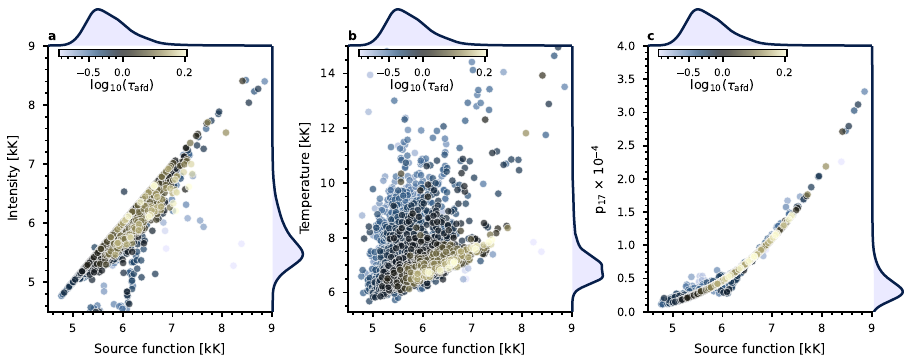}
    \caption{Correlations between the properties of \Hepsilon\ emission profiles and atmospheric quantities. The atmospheric properties have been weighted by the relative contribution functions.
    \emph{Panel a:} Source function and line core intensity (derived from a Gaussian fit) in brightness temperature. \emph{Panel b:} Source function versus gas temperature. \emph{Panel c:} Source function and transition probability $p_{17}$. The color of the points indicates if \Hepsilon\ is optically thin or thick, as shown by the color bar that shows $\log_{10}(\tau_\mathrm{afd})$. The probability density functions of the points, calculated using Gaussian kernel density estimation, are shown at the top and right of each panel.}
    \label{fig:correlation_sf_int}
\end{figure*}

We started with the line core intensity, which we show in Fig.~\ref{fig:correlation_sf_int}.
In general, the core intensity of an optically thick line is set by its source function in the line formation region.
From Fig.~\ref{fig:correlation_sf_int}, we see that indeed the \Hepsilon\ line core intensity exhibits a reasonable correlation with the relative contribution function weighted \Hepsilon\ source function, as demonstrated in panel (a).
The \Hepsilon\ core intensity is a reliable indicator of the source function, but to which parameter is the source function related?
We explore this in the other two panels of Fig.~\ref{fig:correlation_sf_int}.
A natural assumption would be temperature, displayed in panel (b).
However, no significant correlation exists between temperature and the source function, except in the most optically thick regions, where the coupling to temperature is stronger.
In many of the cases, the source function (and therefore line intensity) is decoupled and often below the temperature due to scattering and interlocking, similar to other chromospheric lines \citep[e.g., \MgIIHandK; see][]{Leenaarts2013}.
The parameter that shows a very good correlation with the source function is the transition probability $p_{17}$, shown in panel (c) (The correlation is a curved line because we are plotting the source function in a brightness temperature scale).
Therefore, the \Hepsilon\ line core intensity serves as a proxy for the strength of the transition probability $p_{17}$, which is consistent with what we found for the individual energetic events in Sects. \ref{sec:Ellerman bomb} - \ref{sec: Small-scale flares}.

Another key property of \Hepsilon\ is its width. The widths of Balmer lines are primarily determined by Doppler broadening or linear Stark broadening (from electrons and ions).
\emph{RH 1.5D} includes linear Stark broadening by electrons, following \citet{Sutton1978}, with a dependence on electron density, proportional to $n_e^{2/3}$.
However, we find no correlation between electron density and the width of \Hepsilon\ (not shown in the figures), which indicates that the dominant broadening mechanisms are thermal, nonthermal, and opacity broadening.

\begin{figure*}
    \centering
    \includegraphics[width=1\textwidth]{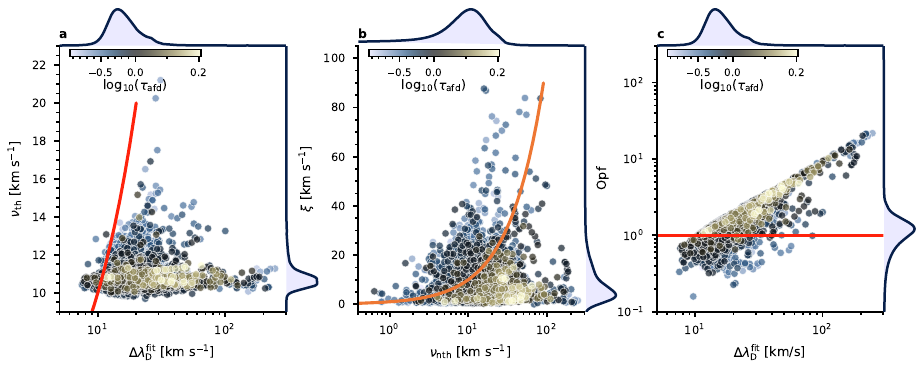}
    \caption{
    Correlations between different contributions to the \Hepsilon\ broadening.
    \emph{Panel a:} Width of \Hepsilon\ ($\Delta \lambda_\mathrm{D}^\mathrm{fit}$, from a Gaussian fit) and thermal broadening ($v_\mathrm{th}$). \emph{Panel b:} Nonthermal broadening ($v_\mathrm{nth}$) and nonthermal velocity ($\xi$). \emph{Panel c:} Width of \Hepsilon\ ($\Delta \lambda_\mathrm{D}^\mathrm{fit}$) and the opacity broadening factor ($Opf$). The color of the points indicates if \Hepsilon\ is optically thin or thick, as shown by the color bar that shows $\log_{10}(\tau_\mathrm{afd})$. The probability density functions of the points, calculated using Gaussian kernel density estimation, are shown at the top and right of each panel. The solid lines in panels \emph{a} and \emph{b} denote the $y=x$ curve.}
    \label{fig:correlation_broadening}
\end{figure*}

In Fig.~\ref{fig:correlation_broadening} we explore the relations between the measured \Hepsilon\ width, $\Delta\lambda_\mathrm{D}^\mathrm{fit}$, the thermal and nonthermal broadening, and the opacity broadening factor.
In panel (a), we compare the thermal broadening $v_\mathrm{th}$, computed using the temperature weighted by the relative contribution function, with the line width.
We find that the thermal broadening alone cannot explain the measured widths, which can be much larger.
Most points fall to the right of the red line, which represents the width expected from thermal broadening alone.
Notably, many regions with larger optical depth show widths many times larger than the thermal broadening.

Since thermal broadening fails to explain the observed widths, we look into how the remaining nonthermal broadening $v_\mathrm{nth}$ from the measured widths can be explained by $\xi$, the nonthermal broadening estimated from velocities in the simulation.
We do this in panel (b) of Fig.~\ref{fig:correlation_broadening}.
The red line in the panel is $v_\mathrm{nth}=\xi$; it delineates excessive (left side) and insufficient (right side) contributions of nonthermal velocity to the line width.
The values of $\xi$ are clustered around $5$~km/s, while $v_\mathrm{nth}$ estimated from the spectral widths is typically larger.
This suggests that a considerable portion of these widths cannot be explained by either thermal or nonthermal mechanisms alone.
This is especially true for locations that are optically thicker, where opacity broadening is likely important.

To quantify the contribution of opacity broadening, we plot in panel (c) the opacity broadening factor (Eq. \ref{eq:opf}) against the theoretical line width (Eq. \ref{eq:theoretical_broadening}).
This relation emphasizes that opacity broadening is significant, particularly for the most optically thick regions.
Points above 1 (the red line) indicate cases where opacity broadening enhances the line width, while points below 1 suggest that the theoretical line width overestimates the observed width.

\begin{figure}
    \centering
    \includegraphics[width=0.5\textwidth]{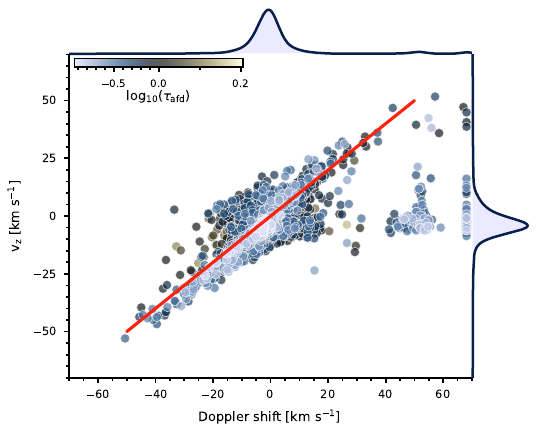}
    \caption{Correlation between line-of-sight velocity, $v_z$, and the fitted Doppler shift of the \Hepsilon\ line core. $v_z$ is weighted by the relative contribution function. The color of the points indicates if \Hepsilon\ is optically thin or thick, as shown by the color bar that shows $\log_{10}(\tau_\mathrm{afd})$. The probability density functions of the points, calculated using Gaussian kernel density estimation, are shown at the top and right.}
    \label{fig:correlation_vel}
\end{figure}

The next line property we looked into was the Doppler shift of \Hepsilon, which we expect can be used to diagnose velocities in the line-forming region.
In Fig.~\ref{fig:correlation_vel} we plot the Doppler shift estimated from the spectra against the simulation's line-of-sight (vertical) velocity weighted by the relative contribution function of \Hepsilon.
The figure illustrates a strong correlation between atmospheric velocities and the Doppler shifts of \Hepsilon.
Optically thin points (light blue) exhibit a clearer velocity-Doppler shift relationship due to their nearly ideal Gaussian profiles.
In contrast, line profiles from optically thick regions often exhibit non-Gaussian shapes (e.g., central reversals), which will affect the quality of the Gaussian fit and therefore the correlation.
Outliers around Doppler shifts of $50$~km/s and $70$~km/s likely result from profiles with central reversals, where the fitting routine misidentifies the line centroid, favoring the shorter wavelength (upflow) emission peak, creating two distinct clusters.

\subsection{\Hepsilon\ and \Halpha}
\label{sec: Hepsilon and Halpha}

The spatial extent and the flame-like structure of Ellerman bombs are best observed at off-vertical inclinations \citep{Rouppe2016}, with a $\mu\equiv\cos\theta=0.5$ being a good compromise between an inclined view but away from the limb proper, where too much superposition occurs.
We generated \Halpha\ and \Hepsilon\ synthetic spectra for different $\mu$ using the \emph{Muspel} package, using the hydrogen populations from \emph{Multi3D} for \Halpha\ and \emph{RH 1.5D} for \Hepsilon. 

For the inclined profiles of \Hepsilon\ we assumed no contribution from \CaH, only \Hepsilon.
This is a strong simplification, but a good approximation for the \Hepsilon\ line core intensity, which is our main goal.
Using results from \emph{RH 1.5D} we find that neglecting \CaH\ has only a minor impact on the \Hepsilon\ line core, and therefore use the \emph{Muspel} \Hepsilon-only synthesis as a good proxy.

\begin{figure*}
    \centering
    \includegraphics[width=1\textwidth]{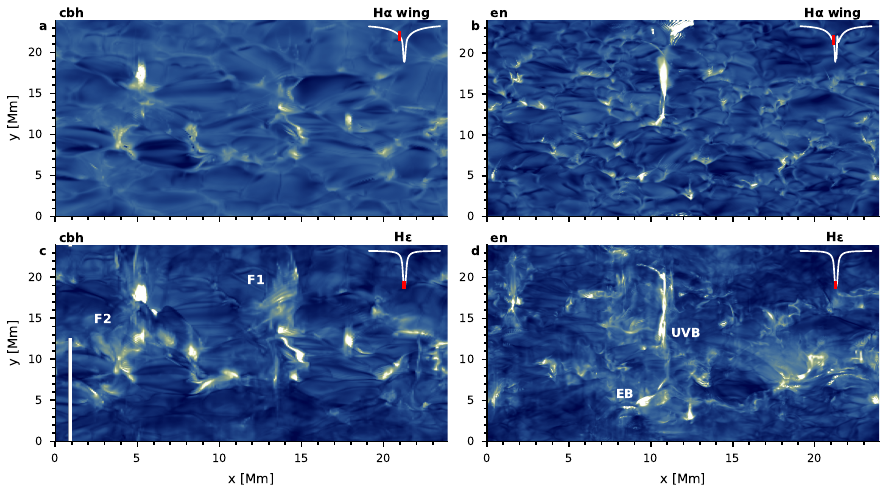}
    \caption{Synthetic \Halpha\ and \Hepsilon\ images at an heliocentric angle of $\mu=0.5$. \emph{Panels a} and \emph{b} show an \Halpha\ wing image for the \texttt{cbh} and \texttt{en} simulations. \emph{Panels c} and \emph{d} show an \Hepsilon\ line core image for the \texttt{cbh} and \texttt{en} simulations. The wavelength position of each image is indicated by a red vertical line on top of the averaged spectral profile at the top-right corner. The \Hepsilon\ images are a proxy for the line core emission and were calculated by neglecting the contribution of \CaH.}
    \label{fig:cb_eb_polar60}
\end{figure*}

We show an inclined view from the simulations at $\mu=0.5$ in Fig. \ref{fig:cb_eb_polar60}. 
The \texttt{cbh} simulation shows multiple brightenings in the \Halpha\ wing image, colocated with \Hepsilon\ line core brightenings.
These brightenings are more intense in \Hepsilon, due to its higher extinction compared to the \Halpha\ wing.
The higher extinction of \Hepsilon\ allows us to observe finer details of these structures, revealing those that could be optically thin in the \Halpha\ wing.
Structures that are otherwise faint in \Halpha\ wing become prominent in \Hepsilon.
This is exemplified by the flare structures F1 and F2, labeled in panel (c), where F1 appears as a tower-like structure with a blurry, veil-like appearance in \Hepsilon, but shows little to no signature in the wing of \Halpha.
The blurry appearance of F1 in \Hepsilon\ results from parts of the flare being optically thin to \Hepsilon\ radiation, an effect even more pronounced for the more energetic flare F2.

The \texttt{cbh} and \texttt{en} simulations contain multiple Ellerman bombs, visible as flame-like structures in the \Halpha\ wing.
For instance, the Ellerman bomb examined in Sect. \ref{sec:Ellerman bomb} appears as a dome-like structure (labeled EB in panel (d)) in both \Halpha\ and \Hepsilon, although the structure is more pronounced and extended in \Hepsilon.
Another notable example is the inverted Y-shaped structure, located around $x \approx 1.5$ and $y \approx 16$~Mm.
This structure is clearly visible in the \Hepsilon\ line, whereas \Halpha\ only shows a faint indication of it.
In general, wherever Ellerman bombs are detected in \Halpha, a co-spatial signature is present in \Hepsilon.
However, \Hepsilon\ reveals more flame-like structures that are not visible in \Halpha. This difference comes from the larger line extinction in \Hepsilon\ compared to the \Halpha\ wing, which enhances the visibility of small-scale structures. 

\section{Discussion}
\label{sec:Discussion}

We investigated the formation and observational properties of \Hepsilon\ for small-scale energetic events, including Ellerman bombs, UV bursts, and small-scale flares.
Our focus lies on understanding how \Hepsilon\ behaves under varying atmospheric conditions connected to energetic events and to what degree these events can be diagnosed with \Hepsilon\ spectra.
Ellerman bombs, UV bursts, and flares each display distinct characteristics in the \Hepsilon\ line.
Due to high temperatures at lower column masses (chromospheric regions), UV bursts and flares appear partially optically thin in \Hepsilon, resulting in enhanced but diffuse structural features.
In contrast, Ellerman bombs form in optically thick conditions with temperature increases occurring at higher column masses, which yields sharper, better-defined structures than those of UV bursts and flares.

However, it is not easy to quantify the temperature increases from \Hepsilon\ spectra.
Its core intensity is primarily determined by the transition probability $p_{17}$, and not by the temperature directly, which one might expect given that small-scale energetic events typically cause temperature enhancements.
The temperature enhancements increase $p_{17}$, producing more photons emitted from small-scale energetic events.

The mechanism of photon creation in \Hepsilon\ differs from the traditional view that the Balmer continuum primarily generates the photons for Balmer line transitions \citep{Gebbie1974, Ayres1975}.
The reason is the following.
The Balmer continuum drives the radiative rates in the hydrogen atom, which leads the atomic transitions to adjust \citep{Carlsson2002}.
The changes in transition rates, particularly within the Lyman series, influence  $p_{17}$, which results in more photons being emitted through the ground state of hydrogen rather than through transitions into the continuum.
This approach contrasts a view where the Balmer continuum is thought to ionize hydrogen, leading to recombination and cascading downward to the upper levels of the Balmer series transitions, which will eventually emit photons.
However, this ionization-recombination model is unsupported by the multilevel source function, which defines the probabilities of transition paths through the atom leading to the population of the upper level of a transition eventually emitting photons \citepalias{Krikova2024}.
This highlights that the emission of photons follows primarily from transitions connected to the Lyman series rather than the ionization-recombination view.
They are connected to temperature in a nonlinear way, and therefore it is difficult to convert from \Hepsilon\ line intensity into temperature.

To date, no high-resolution observations of flares in \Hepsilon\ have been published.
The most comprehensive results available are low-resolution observations of flares by \cite{Rolli1995} and \citet{Rolli1998a, Rolli1998b}, which used the \Hepsilon\ line width to estimate the electron density evolution in the chromosphere during flare events.
These studies assumed optically thin \Hepsilon\ formation, no nonthermal broadening, thermal broadening at  $T = 10^4$~K, and linear Stark broadening to approximate electron density.
However, our analysis of the \Hepsilon\  widths suggests that nonthermal and opacity broadening can substantially impact the line width.
They must be taken into account to obtain accurate estimates of electron density during flares.
Generally, drawing definitive conclusions regarding which broadening mechanisms and atmospheric parameters primarily set the \Hepsilon\ width is challenging, as the full extent of the \Hepsilon\ profile remains ``concealed'' from observers.
The atmosphere is optically thick to \Hepsilon\ wing photons, with \CaH\ extinction dominating and preventing these wing photons from escaping.
As a result, \Hepsilon\ photons can escape only near the line core, which is the visible part in observations.
The more dominant \Hepsilon\ extinction becomes over \CaH\ extinction, which is strongly temperature-dependent (due to \Lyalpha), the more of the \Hepsilon\ profile is revealed.
This explains why certain data points in Fig. \ref{fig:correlation_broadening} exhibit stronger broadening than theoretical predictions (e.g., points to the left of the red line), complicating efforts to find the dominant broadening mechanism.

The Doppler shift of \Hepsilon\ is a good estimate of atmospheric velocities, as shown by its correlation with line-of-sight velocities.
\Hepsilon\ can be used to measure outflows in Ellerman bombs — a measurement that is difficult to perform in \Halpha\ or \Hbeta.
Additionally, \Hepsilon\ could enable detailed observations of atmospheric velocity responses to UV bursts and flares in high-resolution observations, at the atmospheric layers where the \Hepsilon\ line core is formed.
This allows measurements of chromospheric velocities throughout the flare's progression and across different regions of flares.
The velocity estimate works better for Gaussian-shaped \Hepsilon\ profiles, which usually come from optically thin regions.
For optically thicker regions, as seen in Fig. \ref{fig:correlation_vel}, there is more scatter in the correlation because it is difficult to reliably extract shifts from non-Gaussian profiles with central reversals and/or strong asymmetries.

Figure \ref{fig:cb_eb_polar60} shows more Ellerman bomb structures in \Hepsilon\ than in \Halpha.
This is consistent with the observations of \citet{Rouppe2024} and underscores the value of higher-order Balmer series lines to capture details of small-scale energetic events.
For example, the Ellerman bomb structure highlighted in Fig. \ref{fig:cb_eb_polar60} and the inverted Y-shaped structure around $x \approx 1.5$ and $y \approx 16$~Mm are clearly visible in \Hepsilon\ but absent from \Halpha.
These differences arise because the \Hepsilon\ line core has more extinction than the \Halpha\ wing, where such structures are fainter.
If the extinction in small-scale energetic events is amplified due to a temperature or density increase, they can become more visible in the \Halpha\ wing.
%
%
A critical assumption in synthesizing the inclined view of \Hepsilon\ was neglecting \CaH, but we find it a reasonable approximation close to the line core because both the \Hepsilon\ extinction and source function dominate over \CaH.
As a result, our maps show granulation in the \Hepsilon\ core (similar to the \Halpha\ wing) instead of reversed granulation, except in regions of small-scale energetic events.
The bright, diffuse structures in panel (d) reflect contributions from optically thin areas within \Hepsilon.
Furthermore, we are limited by the assumption of 1.5D, but our findings in \citetalias{Krikova2023} indicate that 3D effects are not pronounced for \Hepsilon.

\section{Conclusions}
\label{sec:Conclusion}

We studied the formation of \Hepsilon\ in small-scale energetic phenomena from two active-Sun \emph{Bifrost} simulations.
We find distinct \Hepsilon\ signatures of Ellerman bombs, UV bursts, and flares that can offer additional diagnostics compared to other spectral lines.
Key findings from our study include:
\begin{itemize}
    \item Flares and UV bursts exhibit partial optical thinness to \Hepsilon\ radiation due to strong temperature increases in the chromosphere and should appear as blurred structures in observations.
    \item Ellerman bombs remain largely optically thick in \Hepsilon\ and appear as sharply defined structures, tracing temperature increases in the lower atmosphere.
    \item Ellerman bombs are easier to see in \Hepsilon\ than in \Halpha\ because the line core of \Hepsilon\ has more extinction than the wing of \Halpha.
    \item \Hepsilon\ provides good velocity diagnostics for small-scale heating events in the lower chromosphere.
    \item The amount of \Hepsilon\ emission is a poor tracer of temperature since it is mostly set by the $p_{17}$ transition probability, which depends on Lyman transitions and nonlinearly on temperature.
    \item The width of \Hepsilon\ is poorly correlated with atmospheric properties because it is influenced by various broadening mechanisms: thermal, nonthermal, and opacity broadening.
\end{itemize}

In \citetalias{Krikova2023} we demonstrate that interlocking strongly influences the line source function and, consequently, the amount of emitted \Hepsilon\ photons.
Here, we pinpoint the $p_{17}$ transition as responsible for setting \Hepsilon\ emission in energetic events.
We show that \Hepsilon\ serves as a valuable diagnostic tool for detecting and analyzing small-scale energetic events, such as Ellerman bombs, UV bursts, and small-scale flares, and demonstrate how these phenomena can be observed in \Hepsilon.

Compared to other hydrogen lines, \Hepsilon\ is weaker but has the advantage of not being opaque enough to the fibril canopy. Therefore, it can be used to image lower chromospheric heating in its line core, not just in the wings like \Halpha\ and \Hbeta.
This makes it more reliable for extracting velocities as the line core is easier to trace than wing emission.
Its position in the \CaH\ wing is both a curse and a blessing. The dominating background obscures nearly the whole \Hepsilon\ profile except for a small region around the core.
Other lines such as \Hdelta\ and \Hgamma\ could be promising alternatives, particularly for quiet-Sun Ellerman bombs, if the chromosphere is optically thin in these lines --- a possibility that merits further investigation.

\begin{acknowledgements}
This work has been supported by the Research Council of Norway through its Centers of Excellence scheme, project number 262622. 
Computational resources have been provided by Sigma2 – the National Infrastructure for High-Performance Computing and Data Storage in Norway. KK acknowledges support by the European Research Council under ERC Synergy grant agreement No. 810218 (Whole Sun).
\end{acknowledgements}

\bibliographystyle{aa}
\bibliography{mybib}

\end{document}